\documentclass[journal,transmag]{IEEEtran}

\usepackage{graphicx}
\usepackage{algorithm}
\usepackage{algorithmic}

\usepackage[compress]{cite}
\usepackage{stmaryrd}
\usepackage{mathrsfs}
\usepackage{amsfonts}
\usepackage{amsmath}
\usepackage{cases}
\usepackage{txfonts}
\usepackage{amssymb}
\usepackage{epsfig}
\usepackage{bm}
\usepackage{xcolor}
\usepackage{subcaption}
\usepackage[justification=centering]{caption}

\usepackage{epstopdf}

\begin{document}

\title{Union Bound Analysis for Spin-Torque Transfer Magnetic Random Access Memory (STT-MRAM) with Channel Quantization}

\author{Xingwei Zhong, Kui Cai, \emph{Senior Member, IEEE}, and Guanghui Song  \vspace{-0.4cm}}

\author{\IEEEauthorblockN{Xingwei Zhong\IEEEauthorrefmark{1},
Kui Cai\IEEEauthorrefmark{1},~\IEEEmembership{Senior Member,~IEEE}, and
Guanghui Song\IEEEauthorrefmark{1}}
\IEEEauthorblockA{\IEEEauthorrefmark{1} Singapore University of Technology and Design (SUTD), Singapore, 487372}
}
\IEEEtitleabstractindextext{
\begin{abstract}
As an emerging non-volatile memory (NVM) technology, spin-torque transfer magnetic random access memory (STT-MRAM) has received great attention in recent years since it combines the features of low switching energy, fast write/read speed, and high scalability. However, process variation and thermal fluctuation severely affect the data integrity of STT-MRAM, resulting in both write errors and read errors. Therefore, effective error correction codes (ECCs) are necessary for correcting memory cell errors. Meanwhile, the design of channel quantizer plays a critical role in supporting error correction coding for STT-MRAM. In this work, we propose a union bound analysis which can accurately predict the word error rates (WERs) of ECCs with maximum-likelihood (ML) decoding over the quantized STT-MRAM channel. The derived bound provides a theoretical tool for comparing the performance of ECCs with different quantization schemes at very low error rate levels without resorting to lengthy computer simulations. Moreover, we also propose a new criterion to design the channel quantizer by minimizing the WERs of ECC decoding that are obtained from the union bound analysis. Numerical results show that the proposed union-bound-optimized {(UBO)} quantizer can achieve better error rate performance than the state-of-art quantizers for STT-MRAM.
\end{abstract}
}

\maketitle
\IEEEdisplaynontitleabstractindextext
\IEEEpeerreviewmaketitle

\section{Introduction}
Owing to its superior features of fast write/read speed, low switching energy, and high scalability, spin-torque transfer magnetic random access memory (STT-MRAM) has shown a high potential for the applications as embedded non-volatile memory (NVM) or storage class memory (SCM). However, process variation and thermal fluctuation have a detrimental effect on the data recovery of STT-MRAM, leading to both the read errors and write errors \cite{caij}. Therefore, error correction codes (ECCs) have been employed to improve the data integrity of STT-MRAM \cite{caij,caic}. For example, a (71,64) Hamming code with single-error-correction capability is adopted by Everspin's 16 Mb MRAM. Double-error-correcting Bose-Chaudhuri-Hoquenghem (BCH) codes are also proposed to improve the reliability of STT-MRAM \cite{caic}. Furthermore, the (72,64) extended Hamming code with a hybrid decoding algorithm \cite{caij} has also been proposed for STT-MRAM.


Meanwhile, the channel quantizer that quantizes the signal read back from the STT-MRAM cell is critical to support the above ECCs, since high precision analog-to-digital converters (ADCs) are not applicable for high-speed memories such as STT-MRAM. Various criteria have been proposed for designing the quantizer for STT-MRAM, including the Maximizing-Mutual-Information (MMI) \cite{Kurkoski} criterion, the Maximizing-Cutoff-Rate (MCR) criterion, and the optimizing Polyanskiy-Poor-Verdu-Bound (PPVB) criterion \cite{meic}. However, the final choice of the quantization scheme is still relying on the decoding error rate performance of ECCs. It is typically evaluated by using computer simulations, which are too slow to reach the very low target error rate levels required by the data storage systems\cite{Nowak}.

The union bound analysis is a theoretical performance estimation technique that is used for bounding the error rate of maximum-likelihood (ML) decoding of ECCs. In the literature, it was mainly developed for codes over symmetric channels \cite{Ryan}. In \cite{songc}, more sophisticated union bound techniques were proposed for simple asymmetric channels, {\it i.e.} the Z-channel and the general binary asymmetric channel (BAC). Up till now, no work has been reported for the union bound analysis for the STT-MRAM channel.

In this work, we consider the quantized STT-MRAM channel model with both the write errors and read errors, which can be modeled as a concatenation of a BAC and a binary-input multiple-output discrete memoryless channel (BIMO-DMC). Our proposed union bound analysis first utilizes a function of a predefined multi-dimensional distance between two codewords to calculate the pairwise error probability (PEP) \cite{songc} of the quantized STT-MRAM coding channel. Note that this is different from the cases with the symmetric channels where the PEP is uniquely determined by the Hamming distance of the codewords. Based on the PEP, we derive a union bound of the word error rate (WER) of ML decoding as a function of the weight distributions of ECCs. Next, we further propose to use the WER obtained from the above union bound analysis as the criterion to design the channel quantizer of STT-MRAM. We also develop effective differential evolution (DE) algorithm \cite{Kara} to determine the optimum quantization boundaries for the multi-bit quantizers. Since our proposed union-bound-optimized {(UBO)} quantizer takes into consideration the weight spectrum of ECCs, it can achieve better error rate performance than the prior-art quantizers \cite{Kurkoski,meic}.

The rest of the paper is organized as below. Section II introduces the model of ECC coded STT-MRAM system with channel quantization. Section III presents the union bound analysis for the quantized STT-MRAM channel. Section IV further proposes the {UBO} quantization scheme. The simulation results are showed in Section V. Finally, Section VI concludes the paper.

\section{System Model}
\subsection{Memory Cell Errors of STT-MRAM}
As is illustrated by Fig. 1, an STT-MRAM cell typically has a magnetic tunneling junction (MTJ) as the data storage element and an n-type metal-oxide-semiconductor (nMOS) transistor
as the access control device \cite{Zhangc}. The MTJ consists of a tunneling oxide layer that is sandwiched between two ferromagnetic layers, {\it i.e.} a free layer and a reference layer. While the magnetization direction of the free layer can be switched by changing the direction of the write current, the magnetization direction of the reference layer is fixed. When the direction of the free layer is the same as that of the reference layer, the MTJ is in a low resistance state, which can represent an input information bit of `0'. On the other hand, if the relative direction between the free layer and the reference layer are opposite, the MTJ is in a high resistance state, representing an input information bit of `1'.

\begin{figure}[b]
\centering
\includegraphics[height=1.4in,width=1.1in]{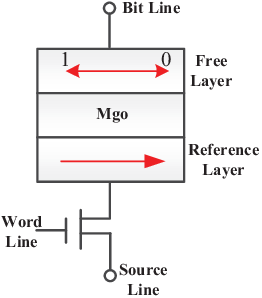}
\caption{The structure of an STT-MRAM cell.}
\label{quantizated_channel_model_full}
\end{figure}

Process variation and thermal fluctuation \cite{Lij,Zhangc} are two major factors that affect the reliability of the data stored in the STT-MRAM cell, resulting in both the write errors and read errors \cite{caij}. In particular, they may cause the write error when the memory cell fails to be switched from one resistance state to the other\cite{Chendynamic}. Furthermore, it has been widely found that the write error rate of $1\rightarrow 0$ switching, denoted by $P_{1\rightarrow 0}$, is much lower than that of the $0\rightarrow 1$ switching, denoted by $P_{0\rightarrow 1}$ \cite{Chendynamic}. The read errors can be classified into the read disturb error (denoted by $P_{rd}$) and the read decision error \cite{Lij}. The read disturb error is caused by an accidental switching of the state of the MTJ. It is also asymmetric which can only occur in one direction depending on the direction of the read current. On the other hand, the read decision error occurs when the two resistance states cannot be differentiated due to widened distributions of the MTJ resistances caused by process variation \cite{Lij}.

\subsection{ECC Coded STT-MRAM System with Channel Quantization}
Based on the major error characteristics of STT-MRAM, a cascaded channel model is proposed in \cite{caij}, which consists of two parts: a BAC and a Gaussian mixture channel (GMC). For reading with write-1 direction, the crossover probabilities of BAC, which model the write errors and the read disturb errors, can be expressed as: $p_{0}=\dfrac{P_{1\rightarrow 0}}{2}+(1-\dfrac{P_{1\rightarrow 0}}{2})P_{rd}$, $p_{1}=\dfrac{P_{0\rightarrow 1}}{2}(1-P_{rd})$, $q_{0}=(1-\dfrac{P_{1\rightarrow 0}}{2})(1-P_{rd})$, $q_{1}=(1-\dfrac{P_{0\rightarrow 1}}{2})+\dfrac{P_{0\rightarrow 1}}{2}P_{rd}$. Here, $p_{0}$ and $p_{1}$ are the probabilities of making a `0' to `1' and making a `1' to `0', respectively, and $q_{0}$ and $q_{1}$ are the probabilities that no error occurs when transmitting the `0' and `1', respectively. Moreover, a GMC is utilized to model the read decision error . The means and variances of the high and low resistances are denoted by $\mu_{1}$, $\sigma_{1}$, $\mu_{0}$ and $\sigma_{0}$, respectively.

In this work, we consider an ECC coded STT-MRAM system with channel quantization. As shown by Fig. 2, the input user data is first encoded by an ECC encoder. The ECC codeword of length $N$ generated, denoted by $\bm{x} = (x_1,\ldots,x_N) \in C$ with $C$ being the ECC code-book, is then transmitted over a quantized STT-MRAM channel which is derived based on the above cascaded channel model \cite{meic}. That is, through quantization, the GMC becomes a BIMO-DMC. In particular, a $q$-bit quantizer maps the GMC output $y_k$, $k\in\{1, \ldots, N\}$, into $M=2^{q}$ outputs $\bar{y}_{k}^{0}, \bar{y}_{k}^{1}, \ldots, \bar{y}_{k}^{M-1}$. The boundaries of quantization intervals are denoted as $\bm{b}=({b}_{0}, \ldots, {b}_{M})$, with ${b}_{0} = -\infty$ and ${b}_{M} = +\infty$. Let $A_{j} = (b_{j},b_{j+1})$ represent the $j$-th quantization interval, $j = 0, 1, \ldots, M-1$. The transition probability of the quantized channel can be derived as \cite{meic}:


\begin{equation}
T(\bar{y}_{k}^{j}|x_{k}=0) = q_{0}Pr(\bar{y}_{k}^{j}|\bar{x}_{k}=0)+p_{0}Pr(\bar{y}_{k}^{j}|\bar{x}_{k}=1),
\label{tran0}
\end{equation}
\begin{equation}
T(\bar{y}_{k}^{j}|x_{k}=1) = p_{1}Pr(\bar{y}_{k}^{j}|\bar{x}_{k}=0)+q_{1}Pr(\bar{y}_{k}^{j}|\bar{x}_{k}=1),
\label{tran1}
\end{equation}
where $T(\bar{y}_{k}^{j}|x_{k}=i)$ can be simplified as $t_{ij}$ with $i\in\{0, 1\}$ and $j\in\{0,\ldots,M-1\}$. Here, $Pr(\bar{y}_{k}^{j}|\bar{x}_{k}=i)$ is the transition probability of the BIMO-DMC. As the BIMO-GMC is modelled as a GMC, we can further obtain
\begin{equation}
Pr(\bar{y}_{k}^{j}|\bar{x}_{k}=i) = Q(\dfrac{b_{j}-\mu_{i}}{\sigma_{i}}) - Q(\dfrac{b_{j+1}-\mu_{i}}{\sigma_{i}}),
\label{AGC1}
\end{equation}
where $Q(x) = \dfrac{1}{\sqrt{2\pi}}\int_{x}^{ + \infty }exp(-\dfrac{u^{2}}{2})du $ is the tail distribution function of the standard normal distribution and we set $Q(\dfrac{t_{0}-\mu_{i}}{\sigma_{i}}) = 1 $ and $Q(\dfrac{t_{M}-\mu_{i}}{\sigma_{i}}) = 0 $. Finally, by considering an ML decoder of the ECC, the transmitted codeword can be estimated through the ML decision rule, given by $\bm{\hat{x}} =\arg\max_{\bm{x}\in C}T^{N}(\bm{\bar{y}}|\bm{x})$.

\begin{figure}[t]
\centering
\includegraphics[height=1.5in,width=3.5in]{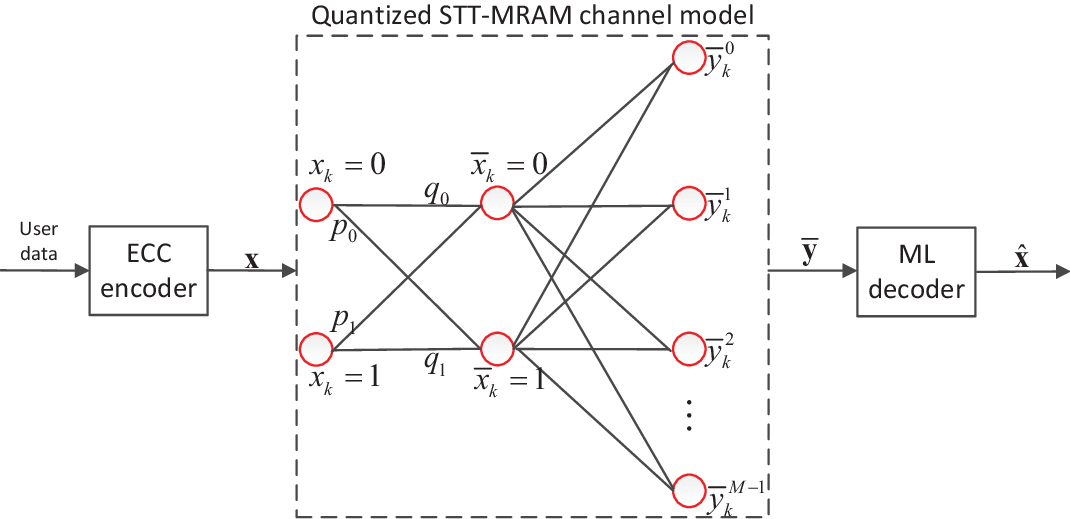}
\caption{ECC coded STT-MRAM system with channel quantization.}
\label{quantizated_channel_model_full}
\end{figure}

%

\section{Union Bound Analysis for the Quantized STT-MRAM Channel}


We first provide some basic definitions that will be used in our subsequent derivations. Let $1\{E\}$ be an indicator function, where $1\{E\}=1$ if event $E$ holds, otherwise, $1\{E\}=0$. For $a, b\in\{0, 1\},c\in\{0,\ldots, M-1\}$, we denote the multi-dimensional distance between vectors $\bm{x}$, $\bm{e}$ and $\bm{\bar{y}}$ as:
\begin{eqnarray}
m_{ab}(\bm{x},\bm{e})=\!\!\!\!\!\!\!\!\!\!\!&&\sum_{k=1}^N\!1\{x_k=a, e_k=b\},\nonumber\\
m_{abc}(\bm{x},\bm{e},\bm{\bar{y}})=\!\!\!\!\!\!\!\!\!\!\!&&\sum_{k=1}^N\!1\{x_k=a, e_k=b, \bar{y}_{k}^{j}=\bar{y}_{k}^{c}\}.\nonumber
\end{eqnarray}
where $\bm{e} = (e_1,\ldots,e_N)$. The above notations can be simplified as  $m_{abc}(\bm{x},\bm{e},\bm{\bar{y}})\buildrel \Delta \over
=m_{abc}$, $m_{ab}(\bm{x},\bm{e})\buildrel \Delta \over =m_{ab}$. Moreover, we can immediately obtain $\sum_{c=0}^{M-1}m_{abc}=m_{ab}$ for $a, b\in\{0, 1\}$.

\subsection{Pairwise Error Probability}
Given two potential quantized channel inputs $\bm{x}, \bm{e}$, PEP $P(\bm{x}\rightarrow \bm{e})$ \cite{songc} is the probability of falsely treating $\bm{x}$ as $\bm{e}$ when $\bm{x}$ is transmitted, and when they are the only two hypotheses. Hence we have
\begin{eqnarray}
P(\bm{x}\rightarrow \bm{e})=\!\!\!\!\!\!\!\!&&\textrm{Pr}\left(T^{N}(\bm{\bar{y}}|\bm{x})\leq T^{N}(\bm{\bar{y}}|\bm{e})| \bm{x}\ \textrm{is sent}\right)\nonumber\\
=\!\!\!\!\!\!\!\!&&\sum_{\bm{\bar{y}}:T^{N}(\bm{\bar{y}}|\bm{x})\leq T^{N}(\bm{\bar{y}}|\bm{e})}T^{N}(\bm{\bar{y}}|\bm{x}).
\label{eq:pep1}
\end{eqnarray}
For the quantized STT-MRAM coding channel with $t_{ij}>0, i\in\{0, 1\}, j\in\{0,\ldots M-1\} $, we have
\begin{eqnarray}
\frac{T^{N}(\bm{\bar{y}}|\bm{e})}{T^{N}(\bm{\bar{y}}|\bm{x})}=\!\!\!\!\!\!\!\!&&\prod_{k=1}^N\frac{T(\bar{y}_{k}^{j}|e_k)}{T(\bar{y}_{k}^{j}|x_k)}\nonumber\\
=\!\!\!\!\!\!\!\!&&\prod_{a,b,c}\left(\frac{t_{bc}}{t_{ac}}\right)^{m_{abc}} = \prod_{c=0}^{M-1}\left(\frac{t_{0c}}{t_{1c}}\right)^{m_{10c}-m_{01c}}. \nonumber
\end{eqnarray}

Based on our definition of the multi-dimensional distance, we can express the error region $E(\bm{x}\rightarrow \bm{e})$, which includes all the possible readback signal vectors that fall into the decision region of $\bm{e}$ although $\bm{x}$ is transmitted, as
\begin{eqnarray}
\!\!\!\!\!\!\!\!\!\!&&E(\bm{x}\rightarrow \bm{e})\nonumber\\
=\!\!\!\!\!\!\!\!\!\!&&\left\{\bm{\bar{y}}\left|\prod_{c=0}^{M-1}\left(\frac{t_{0c}}{t_{1c}}\right)^{m_{10c}-m_{01c}}\geq1, \bm{\bar{y}}\in \{\bar{y}_{k}^{0},\bar{y}_{k}^{1} \ldots \bar{y}_{k}^{M-1}\}\right.\right\}\nonumber\\
=\!\!\!\!\!\!\!\!\!\!&&\left\{\bm{\bar{y}}\left|\sum_{c=0}^{M-1}\left(m_{10c}-m_{01c}\right)\log \left(\frac{t_{0c}}{t_{1c}}\right), \bm{\bar{y}}\in\{\bar{y}_{k}^{0},\bar{y}_{k}^{1} \ldots \bar{y}_{k}^{M-1}\}\right.\right\}.\nonumber
\end{eqnarray}
Then the PEP of (\ref{eq:pep1}) can be calculated as
\begin{eqnarray}
\!\!\!\!\!\!\!\!&&\sum_{\bm{\bar{y}}\in E(\bm{x}\rightarrow \bm{e})}T^{N}(\bm{\bar{y}}|\bm{x})\nonumber\\
=\!\!\!\!\!\!\!\!&& \sum_{E(\bm{x}\rightarrow \bm{e})}\prod_{k=1}^NT(\bar{y}_{k}^{j}|x_k) = \sum_{E(\bm{x}\rightarrow \bm{e})}\prod_{a,b,c}\left(t_{ac}\right)^{m_{abc}}\nonumber\\
=\!\!\!\!\!\!\!\!&&\sum_{\{m_{abc}:\sum_{c=0}^{M-1}\left(m_{10c}-m_{01c}\right) \log \left(\frac{t_{0c}}{t_{1c}}\right)\geq 0, \sum_{c=0}^{M-1}m_{abc}=m_{ab}\}}\nonumber\\
\!\!\!\!\!\!\!\!&& \times \prod_{a,b}\binom{m_{ab}}{m_{ab0},m_{ab1}, \ldots ,m_{ab(M-1)}}\prod_{a,b,c}\left(t_{ac}\right)^{m_{abc}}\nonumber\\
=\!\!\!\!\!\!\!\!&&\sum_{\{m_{abc},a\neq b:\sum_{c=0}^{M-1}\left(m_{10c}-m_{01c}\right) \log \left(\frac{t_{0c}}{t_{1c}}\right)\geq 0, \sum_{c=0}^{M-1}m_{abc}=m_{ab}\}}\nonumber\\
\!\!\!\!\!\!\!\!&& \times \prod_{a\neq b}\binom{m_{ab}}{m_{ab0},m_{ab1}, \ldots ,m_{ab(M-1)}}\prod_{a\neq b,c}\left(t_{ac}\right)^{m_{abc}}\nonumber\\
=\!\!\!\!\!\!\!\!&&\sum_{\{m_{abc}:\sum_{c=0}^{M-1}\left(m_{10c}-m_{01c}\right) \log \left(\frac{t_{0c}}{t_{1c}}\right)\geq 0, \sum_{c=0}^{M-1}m_{abc}=m_{ab}\}}\binom{m_{01}}{m_{010}, \ldots ,m_{01(M-1)}}\nonumber\\
\!\!\!\!\!\!\!\!&&  \times \binom{m_{10}}{m_{100}, \ldots ,m_{10(M-1)}} \prod_{c}\left(t_{0c}\right)^{m_{01c}}\left(t_{1c}\right)^{m_{10c}}.
\end{eqnarray}

Note that due to the fact that the quantized STT-MRAM coding channel is asymmetric, the PEP is a function of the multi-dimensional distances $m_{01}$ and $m_{10}$. For example, when we consider the $2$-bit quantization with $M = 4$, the PEP becomes

\begin{eqnarray}
P(\bm{x} \rightarrow \bm{e})=\!\!\!\!\!\!\!\!&&\sum_{\{m_{abc}:\sum_{c=0}^{3}\left(m_{10c}-m_{01c}\right) \log \left(\frac{t_{0c}}{t_{1c}}\right)\geq 0, \sum_{c=0}^{3}m_{abc}=m_{ab}\}} \nonumber\\
\!\!\!\!\!\!\!\!&&  \times \binom{m_{01}}{m_{010},m_{011},m_{012},m_{013}} \binom{m_{10}}{m_{100},m_{101},m_{102},m_{103}} \nonumber\\
\!\!\!\!\!\!\!\!&& \times \prod_{c=0}^{3}\left(t_{0c}\right)^{m_{01c}}\left(t_{1c}\right)^{m_{10c}}.\nonumber
\end{eqnarray}

\subsection{Union Bound}
When the actual channel input is $\bm{x}$, the decoding WER can be estimated using the union bound as
   \begin{eqnarray}
   p_e\leq\!\!\!\!\!\!\!\!&&\sum_{\bm{e}\neq \bm{x}, \bm{e}\in C}P(\bm{x}\rightarrow \bm{e})\nonumber\\
   =\!\!\!\!\!\!\!\!&&\sum_{d}\sum_{\{\bm{e}\in C,d_H(\bm{x},\bm{e})=d\}}\sum_{m_{01},m_{10}}1\left\{m_{ab}(\bm{x},\bm{e})=m_{ab},a\in\{0, 1\},\right.\nonumber\\
   \!\!\!\!\!\!\!\!&&\ \ \ \ \ \ \ \ \ \ \ \ \ \ \ \ \ \ \ \ \ \ \ \ \ \ \ \    \left.b=1-a\right\}P(\bm{x}\rightarrow \bm{e}).
   \end{eqnarray}
where $d_H(\bm{x},\bm{e})$ is the Hamming distance between $\bm{x}$ and $\bm{e}$. From (6), we know that the union bound can be treated as a function of $m_{ab}$ which depends on the actual channel input. By denoting the number of codewords with Hamming weight $d$ in $C$ to be $A_d, d=1, \ldots, N$, we obtain the following WER bound as
\begin{eqnarray}
p_e\leq\!\!\!\!\!\!\!\!\!\!\!&&\sum_{d}A_d2^{-d}\sum_{m_{01}+m_{10}=d}\binom{d}{m_{01},m_{10}}\nonumber
\end{eqnarray}
\begin{eqnarray}
\!\!\!\!\!\!\!\!&& \times \sum_{\{m_{abc}:\sum_{c=0}^{M-1}\left(m_{10c}-m_{01c}\right) \log \left(\frac{t_{0c}}{t_{1c}}\right)\geq 0, \sum_{c=0}^{M-1}m_{abc}=m_{ab}\}}\binom{m_{01}}{m_{010}, \ldots ,m_{01(M-1)}}\nonumber\\
\!\!\!\!\!\!\!\!&& \times  \binom{m_{10}}{m_{100}, \ldots ,m_{10(M-1)}}  \prod_{c}\left(t_{0c}\right)^{m_{01c}}\left(t_{1c}\right)^{m_{10c}}.\nonumber
\end{eqnarray}

Moreover, by considering the dominant minimum distance, the union bound can be approximated as
\begin{eqnarray}
p_e\approx\!\!\!\!\!\!\!\!\!\!\!&&\sum_{d\in \{ d_{min} \}}\sum_{\{\bm{e}\in C,d_H(\bm{x},\bm{e})=d\}}\sum_{m_{01},m_{10}}1\left\{m_{ab}(\bm{x},\bm{e})=m_{ab},a\in\{0, 1\},\right.\nonumber\\
   \!\!\!\!\!\!\!\!&&\ \ \ \ \ \ \ \ \ \ \ \ \ \ \ \ \ \ \ \ \ \ \ \ \ \ \ \    \left.b=1-a\right\}P(\bm{x}\rightarrow \bm{e})\nonumber\\
 =\!\!\!\!\!\!\!\!&&\sum_{d\in \{ d_{min} \}}A_d2^{-d}\sum_{m_{01}+m_{10}=d}\binom{d}{m_{01},m_{10}}\nonumber\\
\!\!\!\!\!\!\!\!&& \times \sum_{\{m_{abc}:\sum_{c=0}^{M-1}\left(m_{10c}-m_{01c}\right) \log \left(\frac{t_{0c}}{t_{1c}}\right)\geq 0, \sum_{c=0}^{M-1}m_{abc}=m_{ab}\}}\binom{m_{01}}{m_{010}, \ldots ,m_{01(M-1)}}\nonumber\\
\!\!\!\!\!\!\!\!&& \times  \binom{m_{10}}{m_{100}, \ldots ,m_{10(M-1)}}  \prod_{c}\left(t_{0c}\right)^{m_{01c}}\left(t_{1c}\right)^{m_{10c}}.
\label{eq:pepfinal}
\end{eqnarray}
Hence only the code's weight spectrum will affect the above union bound. For example, when we consider the $2$-bit quantization with $M = 4$, the union bound becomes

\begin{eqnarray}
p_e\approx\!\!\!\!\!\!\!\!\!\!\!&&\sum_{d\in \{ d_{min} \}}A_d2^{-d}\sum_{m_{01}+m_{10}=d}\binom{d}{m_{01},m_{10}}\times \nonumber\\
\!\!\!\!\!\!\!\!&& \sum_{\{m_{abc}:\sum_{c=0}^{3}\left(m_{10c}-m_{01c}\right) \log \left(\frac{t_{0c}}{t_{1c}}\right)\geq 0, \sum_{c=0}^{3}m_{abc}=m_{ab}\}}\binom{m_{01}}{m_{010},m_{011},m_{012},m_{013}}\nonumber\\
\!\!\!\!\!\!\!\!&& \times \binom{m_{10}}{m_{100},m_{101},m_{102},m_{103}}\prod_{c=0}^{3}\left(t_{0c}\right)^{m_{01c}}\left(t_{1c}\right)^{m_{10c}}.\nonumber
\end{eqnarray}

\section{Union-Bound-Optimized {(UBO)} Quantization Scheme}
Although various information theoretic analyses based criteria are proposed for designing the channel quantizer for STT-MRAM, such as the MMI criterion, the MCR criterion, and the optimizing PPVB criterion \cite{meic}, none of these criteria have taken into consideration the weight spectrum of the specific ECC applied to the STT-MRAM channel. In this work, we propose a novel criterion to design the channel quantizer for STT-MRAM, by utilizing the union bound analysis derived in the previous section. That is, we propose to choose the quantization boundaries $\bm{b}=({b}_{1}, \ldots, {b}_{M-1})$, by minimizing the WER bound of (\ref{eq:pepfinal}) for the STT-MRAM channel. The obtained quantizer is referred to as the {UBO} quantizer.


In this work, we develop a DE algorithm that is customized based on the work of \cite{Kara} for finding the optimum quantization boundaries. It is summarized by \textbf{Algorithm 1}. In particular, we first set the quantization boundaries $\bm{b}=({b}_{1}, \ldots, {b}_{M-1})$ over the interval of $[\mu_0-4\sigma_0 ,\mu_1+ 4\sigma_1]$ as the input parameter vector of our DE algorithm, and our goal is to find the elements of $\bm{b}$ such that the WER bound of (\ref{eq:pepfinal}) we derived is minimized. Therefore, the cost function of our DE algorithm is defined to be the WER bound of (\ref{eq:pepfinal}). The other initialization parameters of our DE algorithm are set as follows: the population size $N_p=10M$, maximum iteration number $gen_{max}=100$, scaling factor $F=0.8$, crossover rate $CR=0.5$, and the random variable $rand$ follows the uniform distribution between $[0,1]$.

\begin{figure}[t]
\centering
\includegraphics[width=2.5in, height=1.3in]{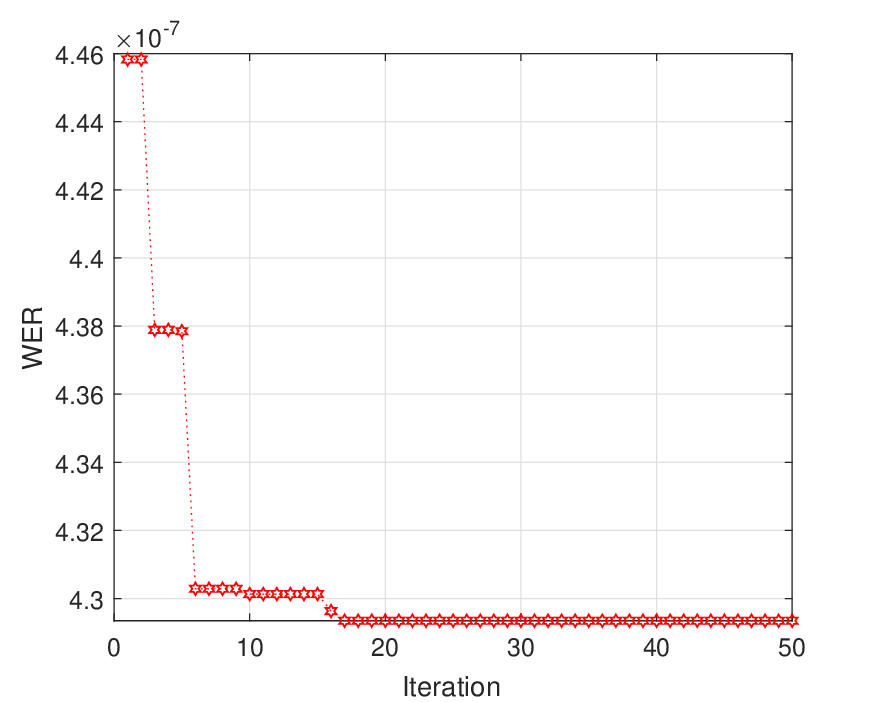}
\caption{Convergences speed of DE for the {UBO} quantizer with $q= 2$, $d_{min}=4$ and $A(d_{min})=8157$ at $\sigma_{0}/\mu_{0} = 9\%$ and $P_1= 1\times 10^{-5}$.}
\label{DE}
\end{figure}

In Figure \ref{DE}, we illustrate the convergence behavior of \textbf{Algorithm
1}. Observe that irrespective of the local optima ({\it i.e.} the plateau) that occurs during the first few iterations, the WER decreases very fast and converges after 17 iterations. This demonstrates the effectiveness of the DE algorithm in searching for
the optimum quantization boundaries.

\begin{algorithm}[t!]
\caption{Searching for the optimum boundaries by DE}
\label{algo: DEUB}
\begin{algorithmic}[1]
\begin{scriptsize}
\REQUIRE $N_p, gen_{max}, F, CR, rand, \bm{b}$.
\ENSURE $\bm{b}$.

\FOR    {($count=0; count<gen_{max}; count++$)}
  \FOR    {($i=0; i<N_p; i++$)}
\STATE  $//$\textit{Mutate/crossover}
    \STATE  $do\  r\leftarrow rand*N_p;\ while (r==i);$
    \STATE  $do\  s\leftarrow rand*N_p;\ while (s==i||s==r);$
    \STATE  $do\  z\leftarrow rand*N_p;\ while (z==i||z==r||z==s);$
    \STATE  $j\leftarrow rand*(M-1);$
    \FOR    {($k=1; k<=(M-1); k++$)}
       \IF{($rand<CR||k==(M-1)$)}
       \STATE  $trial[j]=\bm{b}[z][j]+F*(\bm{b}[r][j]-\bm{b}[s][j]);$
       \ELSE
       \STATE  $trial[j]=\bm{b}[i][j];\ j=(j+1)\%(M-1);$
       \ENDIF
       \ENDFOR
\STATE  $//$\textit{Select, here score and cost are derived by (7)}
       \STATE  $score=P_e(trial);\  cost=P_e(\bm{b});$
       \IF{($score<=cost[i]$)}
       \STATE  $\textbf{for}\ \ (j=1;j<M;j++)\  \bm{b}[i][j]=trial[j];$
       \ELSE
       \STATE  $\textbf{for}\ \ (j=1;j<M;j++)\  \bm{b}[i][j]=\bm{b}[i][j];$
       \ENDIF
  \ENDFOR
\ENDFOR
\RETURN $\bm{b}$.
\end{scriptsize}
\end{algorithmic}
\end{algorithm}

\begin{figure}[t]
\centering
\includegraphics[height=0.49\columnwidth]{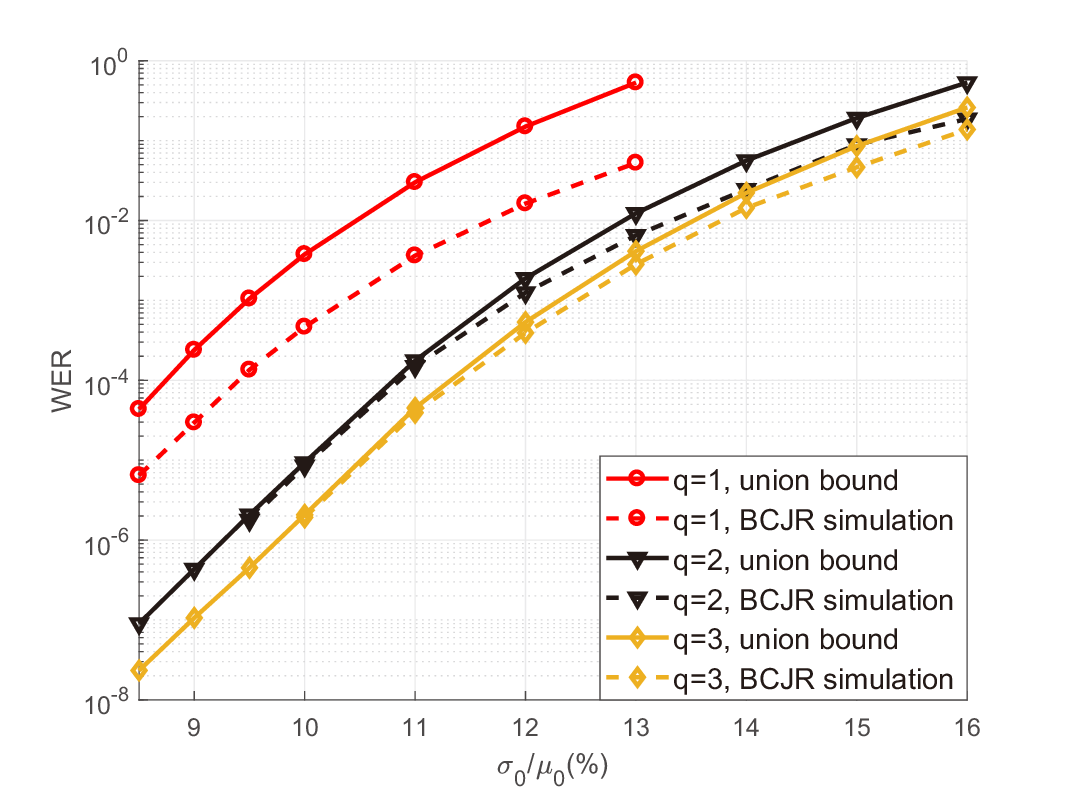}
\caption{Comparison between analytical and simulation WERs, with {UBO} quantizer with $q= 1,2,3$.}
\label{m_bit}
\end{figure}

%

\section{Numerical and Simulation Results}
In our numerical evaluations and computer simulations, following the literature \cite{caij,meic,Zhangc}, we assume that the STT-MRAM cell is with a 45nm$\times$90nm in plane MTJ under a PTM 45nm technology node, with $\mu_{1} = 2k\Omega$, $\mu_{0} = 1k\Omega$, and $\sigma_{0}/\mu_{0} = \sigma_{1}/\mu_{1}$. We assume the write error rate of $P_1= 1\times 10^{-5}$, and vary the mean normalized resistance spreads $\sigma_{0}/\mu_{0}$ (and hence $\sigma_{1}/\mu_{1}$) to account for the influence of different process variations. Moreover, a (72, 64) extended Hamming code with $d_{min}=4$ and $A(d_{min})=8157$ \cite{Davy}, which shows superior error performance over the STT-MRAM channel, is adopted to validate our analysis. The simulated ML decoding performance is obtained using the Bahl, Cocke, Jelinek and Raviv (BCJR) decoder  \cite{Ryan}.

Fig. \ref{m_bit} compares the decoding WERs provided by the union bound with those obtained by the BCJR decoder through simulations, with the proposed {UBO} quantizer with different number of quantization bits $q$. It can be observed that there is a noticeable gap between the union bound and simulated WERs for the case of $q=1$. However, when the number of quantization bits is increased to $q=2$ or more, the performance gap becomes negligible and when the WER is less than $10^{-6}$, the WERs of the union bound are almost identical to the simulated WERs.

\begin{figure}[t]
\centering
\includegraphics[width=3.5in, height=2.6in]{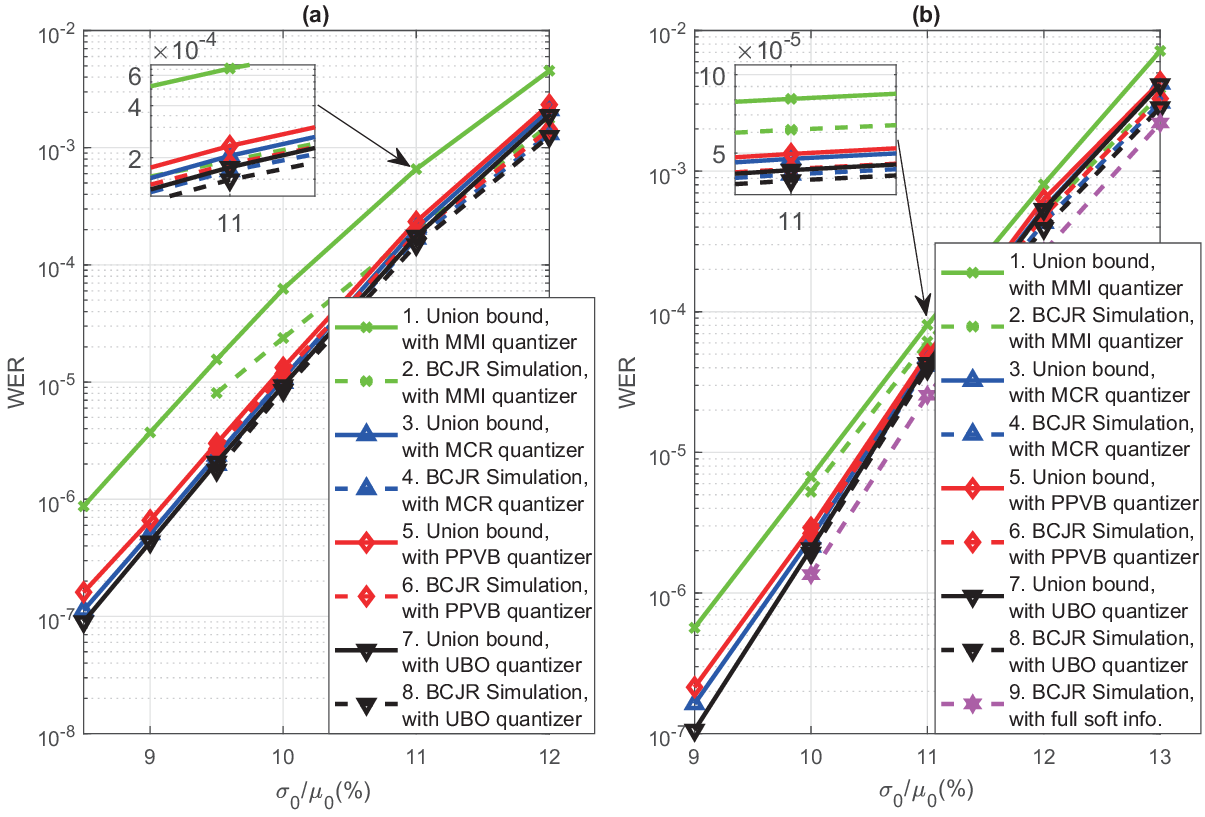}
\caption{Comparison between analytical and simulation WERs with different quantizers, for (a) $q= 2$; (b) $q= 3$ \& $q\rightarrow \infty$.}
\label{2_bit}
\end{figure}

Next, in Fig. \ref{2_bit}(a), we keep $q=2$ and compare the WERs with different types of quantizers, including the MMI, MCR, PPVB, and {UBO} quantizers. A nice agreement is observed again between the WERs predicted by the union bound and those obtained from simulations, for the various types of quantizers. Moreover, our proposed {UBO} quantizer outperforms all the other quantizers, since the corresponding union bound calculation takes into consideration the weight spectrum of the specific (72, 64) extended Hamming code that is applied to the STT-MRAM channel. There is a larger performance improvement over the MMI quantizer, which is most widely adopted in the literature \cite{Kurkoski}. This demonstrates the potential of our proposed union bound analysis in predicting the decoding error rate performance and in guiding the design of the channel quantizer for STT-MRAM.

{Finally, in Fig. \ref{2_bit}(b), we illustrate the WER performance for the case of $q=3$ (Curves 1 to 8). Observe that the trends and relative performances of the different curves associate with different quantizers are similar compared to the case of $q = 2$. Furthermore, the performance improvement for the case of $q=3$ over the case of $q=2$ is quite limited. The reason is because with $q=3$, the WERs of most quantizers (except for the MMI quantizer) are actually approaching those of the BCJR decoder (the ML decoder) with full channel soft information ({\it i.e.} with $q\rightarrow \infty$), which are indicated by Curve 9 of Fig. \ref{2_bit}(b). }

\section{Conclusions}
We have proposed a union bound analysis to accurately predict
the WERs of ECCs applied to the quantized STT-MRAM channel with ML decoding. In particular, we first represented the PEP as a function of a multi-dimensional code distance for the asymmetric quantized STT-MRAM channel. We then summed over different PEPs to get the union bound. We further utilized the WER obtained from the union bound analysis as the criterion to guide the design of the channel quantizer. By applying an effective DE algorithm, the optimum quantization boundaries are determined, and the resulting {UBO} quantizer achieves better WER performance than the state-of-art quantizers for the STT-MRAM channel.

%

\section*{Acknowledgement}
This work is supported by RIE2020 Advanced Manufacturing and Engineering (AME) programmatic grant A18A6b0057 and Singapore MOE Tier 2 fund MOE2016-T2-2-054.



%

\begin{thebibliography}{1}

%
%
%
%
%
%
%
%
%
%
%
%



%
%
%
%
%
%
%
%
%
%
%
%
%
%

\bibitem{caij} K. Cai, and K. A. S. Immink, ``Cascaded channel modeling, analysis, and
hybrid decoding for spin-torque transfer magnetic random access
memory (STT-MRAM)," \emph{IEEE Trans.
Magn.}, vol.~53, no.~11, pp.~1--11, 2017.


\bibitem{caic} K. Cai et al., ``Channel capacity and soft-decision decoding of LDPC codes for spin-torque transfer magnetic random
access memory (STT-MRAM)," in \emph{Proc. Int. Conf. Comput., Netw.,
Commun. (ICNC)}, Jan. 2013, pp.~550--554.

\bibitem{Kurkoski}B. M. Kurkoski et al., ``Quantization of binary-input discrete memoryless channels," \emph{IEEE Trans. Inf. Theory}, vol. 60, no. 8, pp. 4544552, 2014.

\bibitem{meic}Z. Mei et al., ``Information Theoretic Bounds Based Channel Quantization Design for Emerging Memories," in \emph{Proc. IEEE Information Theory Workshop (ITW)}, Nov. 2018, pp.~1--5.

\bibitem{Nowak}J. J. Nowak et al., ``Demonstration of ultralow bit error rates for spintorque
magnetic random-access memory with perpendicular magnetic
anisotropy," \emph{IEEE Magn. Lett.}, vol. 2, pp. 3000204, 2011.

\bibitem{Ryan}L. Bahl et al., ``Optimal decoding of linear codes for minimizing symbol error rate," \emph{IEEE Trans. Inf. Theory}, vol.~20, no.~2, pp.~284--287, 1974.


\bibitem{songc}G. Song et al., ``A union bound analysis for codes over binary asymmetric channels," in \emph{Proc.  IEEE International Conference on Communications (ICC)}, May. 2017, pp.~1--5.

\bibitem{Kara}D. Karaboga and S. Okdem, ``A simple and global optimization algorithm for engineering problems: differential evolution algorithm,"  \emph{Turkish J.
Elect. Eng. Comput. Sci.}, vol.~12, no.~1, pp.~53--60, 2004.

\bibitem{Zhangc}Y. Zhang et al., ``STT-RAM cell design optimization for persistent and non-persistent error rate reduction: A statistical design view," in \emph{Proc. IEEE/ACM Int. Conf. Comput.-Aided Design}, Nov. 2011, pp.~471--477.

\bibitem{Lij}J. Li et al., `` Design paradigm for robust spin-torque transfer magnetic RAM (STT-MRAM) from circuit/architecture perspective,"  \emph{IEEE Trans. Very Large Scale
Integr. Syst.}, vol.~18, no.~12, pp.~1710--1723, 2010.

\bibitem{Chendynamic}B. J. Chen et al., ``A portable dynamic switching model for perpendicular magnetic tunnel junctions
considering both thermal and process variations,"  \emph{IEEE Trans. Magn.}, vol.~51, no.~11, pp.~1300704, 2015.


\bibitem{Davy}A. A. Davydov et al., ``Optimization of shortened hamming codes,"  \emph{Problems of Inform. Transm.}, vol.~17, no.~4, pp.~261--267, 1981.









%
%
%
%
%
%
%
%
%
%
%
%




%
%
%
%
%
%
%



\end{thebibliography}
%
%
%
%

\end{document}